\documentclass{aa}  
\usepackage{xcolor}
\usepackage{graphicx}
\usepackage{txfonts}
\usepackage{lipsum}
\usepackage{subcaption}         % necessary for continued figures, example in section 3
\usepackage{lscape}             % to rotate a single page table, example in appendix.
\usepackage{placeins}           % useful with \FloatBarrier, to keep 
\usepackage[colorlinks,
            linkcolor=red, 
            anchorcolor=green, 
            citecolor=blue %{[RGB]{0,0,128}}, 
            ]{hyperref}
            
\begin{document}
   \title{Large-scale chaos in restricted hierarchical triples driven by short-range forces}   
    \author{Xiaoyan Leng\inst{1,2}
        \and Hanlun Lei\inst{1,2}\fnmsep\thanks{Corresponding author: leihl@nju.edu.cn}
        }
   \institute{
        School of Astronomy and Space Science, Nanjing University, Nanjing 210023, China
        \and Key Laboratory of Modern Astronomy and Astrophysics in Ministry of Education, Nanjing University, Nanjing 210023, China}

   %\date{Received September 30, 20XX}

  \abstract
   {The eccentric von Zeipel–Lidov–Kozai effect, which is widely applied to diverse astrophysical settings, can drive the inner binary to extremely high eccentricities, where short-range forces such as general relativity (GR) become prominent.}
   {Poincar\'{e} surfaces of section show that GR effects reshape phase-space structures, giving rise to a widespread and large-scale chaotic sea. This work aims to uncover the underlying mechanism of GR-enabled chaos.} 
   {The dynamical structures are studied within an adiabatic framework, where an adiabatic invariant is constructed. Phase portraits, defined by the level curves of this invariant, establish analytical boundaries for the chaotic domains. Numerical simulations are subsequently performed to map orbital flips, maximum eccentricities, and Fast Lyapunov Indicators (FLI) across the initial eccentricity–inclination parameter space.}
   {Phase portrait analysis demonstrates that GR precession introduces an uncertainty zone near polar inclinations, where trajectories crossing this zone inherently become chaotic. Leveraging this framework, the chaotic boundaries across the full parameter space are analytically derived, yielding excellent agreement with numerical maps of flipping orbits, maximum eccentricity, and FLI.}
   {In the presence of short-range forces, flipping orbits accompanied by extreme eccentricity excitation are fundamentally chaotic, and the underlying mechanism of large-scale chaos originates from the periodic passage of trajectories through the uncertainty zone in phase space over secular timescales.}
   \maketitle
   \nolinenumbers
%%%%%%%%%%%%%%%%%%%%%%%%%%%%%%%%%%%%%%%%%%%%%%%%%%%%%%%%%%%%
\section{Introduction}
\label{Sect1}

Hierarchical triple systems are ubiquitous across a wide range of astrophysical scales in the Universe, ranging from planetary satellites to supermassive black-hole binaries \citep{naoz2016,shevchenko2016lidov}. In such systems, the separation between the third body and the inner binary is much larger than the orbital dimension of the inner binary, implying that the ratio of their semi-major axes, $\alpha \equiv a/a_{\text{p}}$, is a small parameter. The Hamiltonian can thus be expanded as a power series in $\alpha$ \citep{harrington1968,harrington1969}. To study long-term evolutions, it is common to perform double averaging (DA) over both the inner and outer orbital periods, known as secular (or phase-averaged) approximation \citep{ford2000,naoz2016}.

For circular outer orbits, the leading quadrupole-order perturbation from the outer body excites coupled evolution of eccentricity and inclination if the inclination of the inner orbit relative to the outer plane lies between $39.2^\circ$ and $140.8^\circ$ \citep{vonZeipel1910, lidov1962, kozai1962}. This phenomenon is now commonly referred to as the von Zeipel--Lidov--Kozai (ZLK) effect \citep{ito2019}. When the perturber's orbit is eccentric, the octupole-order term should be incorporated into the Hamiltonian, and the vertical angular momentum of the test particle is no longer conserved \citep{naoz2016}. The resulting long-term modulation of ZLK cycles can then induce repeated prograde--retrograde flips and chaotic behavior. Accordingly, this mechanism is referred to as the eccentric ZLK effect \citep{lithwick2011,ito2019}. The (eccentric) ZLK mechanism has been broadly invoked to explain diverse astrophysical phenomena \citep{naoz2016}, including the formation of hot Jupiters on highly inclined orbits via tidal dissipation \citep{wu2003,fabrycky2007,naoz2011}, as well as facilitating mergers of compact binaries and stellar-mass black holes \citep{thompson2011,liu2017}.

The onset condition and dynamical nature of the eccentric ZLK mechanism have been widely explored. \cite{lithwick2011} systematically explored the eccentric ZLK effect at octupole order via numerical survey, mapping initial conditions leading to orbital flips and extreme eccentricities. Meanwhile, \cite{katz2011} performed a further averaging of the secular equations over the ZLK cycle period, achieving a new constant of motion and an analytical criterion for orbital flips. \cite{li2014a} explored chaos and quasi-periodic evolution in the eccentric ZLK regime using Poincar\'{e} surfaces of section and Lyapunov exponents. Their subsequent study \citep{li2014b} revealed that orbit flips are not limited to low-eccentricity, high-inclination (LeHi) configurations, but can also occur in high-eccentricity, nearly coplanar (HeLi) regimes. The underlying dynamical mechanisms of these two flip channels are fundamentally distinct: LeHi flips are driven by a combination of quadrupole and octupole resonances, whereas HeLi flips are primarily dominated by the octupole resonance. \cite{antognini2015} derived the characteristic timescales of ZLK oscillations at both the quadrupole and octupole levels. Applying the perturbation framework of \cite{henrard1990}, \cite{sidorenko2018} interpreted the eccentric ZLK effect in the LeHi regime as a resonant phenomenon. \cite{lei2022AA} extended this resonance paradigm to the entire parameter space, demonstrating that the eccentric ZLK effect is dynamically equivalent to a three-dimensional apsidal precession resonance in restricted hierarchical planetary systems. Meanwhile, \cite{lei2022AJ} employed three complementary approaches to reconstruct the initial conditions of orbital flips, including Poincar\'e surfaces of section, dynamical system theory (periodic orbits and invariant manifolds), and perturbation theory, showing that flipping orbits are quasi-periodic (or resonant) trajectories around stable, polar periodic orbits.

The classical studies discussed above adopt the DA approximation, which is available if $P_{\rm in} \ll P_{\rm out} \ll t_{\rm ZLK}$, where $P_{\rm in}$ and $P_{\rm out}$ are the inner and outer orbital periods, respectively, and $t_{\rm ZLK}$ is the characteristic ZLK timescale. In addition, since the evolution of DA orbital elements proceeds most rapidly at high eccentricity, the validity of the DA model additionally requires \citep{liu2018,klein2026}
\begin{equation}\label{EqI1}
    t_{\rm ZLK}\sqrt{1 - e_{\max}^2} \gtrsim P_{\rm out}.
\end{equation}
If this condition is violated, the variation in the inner orbital angular momentum over a single outer period becomes non-negligible, causing the DA approximation to break down \citep{liu2018,grishin2018}. Two principal formulations have been developed to overcome this limitation: incorporating the Brown Hamiltonian as a non-linear correction to the DA Hamiltonian \citep{brown1936stellar,soderhjelm1975three,cuk2004secular,breiter2015secular,luo2016,grishin2017generalized,tremaine2023,grishin2024irregular,klein2024hierarchical,lei2025extensionsI,gao2025eccentric}, and adopting the single-averaging (SA) approximation \citep{grishin2018,liu2018,hamilton2019secular,hamilton2024,huang2026,klein2026}.

Even with these corrections, a purely gravitational description may break down when the (eccentric) ZLK effect forces the test particle into extremely close pericenter passages with the central body. Under such extreme conditions, various short-range forces (SRFs) become dominant, such as General Relativistic (GR) precession, tidal interactions, and rotational distortions \citep{wu2003,fabrycky2007,liu2015,hamilton2021secular,naoz2013}. In this regard, \citet{liu2015} systematically evaluated the impact of these short-range forces on the eccentric ZLK effect at octupole order, demonstrating that SRFs impose a firm upper bound on the achievable maximum eccentricity and reduce the inclination parameter space permitting orbital flips. Recently, \citet{huang2026} further analyzed the maximum eccentricity attainable under the joint action of octupole perturbations and GR precession, showing that the limiting eccentricity for flipping orbits can be derived analytically from the quadrupole-order Hamiltonian. By contrast, \citet{hamilton2024} examined test-particle quadrupole dynamics within the SA approximation including GR precession. Focusing on evolution near eccentricity peaks, they identified relativistic phase-space diffusion (RPSD): when the timescale spent near maximum eccentricity becomes comparable to or shorter than the outer orbital period, the binary can jump to an adjacent dynamical trajectory on the outer orbital timescale, thereby breaking the adiabatic invariants of the DA model \citep{hamilton2024,rasskazov2024orbital}. Furthermore, \citet{klein2026} broadened this framework to encompass additional SRFs (e.g., tidal and rotational bulges), showing that all SRFs can induce discrete, non-adiabatic jumps in the binary's effective invariants during high-eccentricity phases, which ultimately catalyze rather than suppress extreme eccentricity excursions. In general, each SRF drives apsidal precession of the argument of pericenter at a rate that scales as $|\dot{\omega}_{\mathrm{SRF}}| \propto (1 - e^2)^{-n}$, where $n = 1$ for GR, $n \simeq 4.5$ for equilibrium tides, and $n = 1.5$ for rotational bulges \citep{klein2026}. For concreteness, we include only GR precession as a representative SRF in this work, while our analysis is applicable to other types of SRFs.

In restricted hierarchical triples, the inclusion of GR precession reshapes the dynamical landscape. By constructing Poincar\'{e} surfaces of section, we find a striking feature: regions of phase space that correspond to regular, ordered motion in the purely Newtonian model become engulfed in a widespread, large-scale chaotic sea once GR precession is included. However, the dynamical mechanism driving this transition to chaos remains unclear. In this work, we systematically investigate the onset of large-scale chaos enabled by short-range forces, with the purpose of uncovering the underlying mechanism.

The remaining structure of this paper is organized as follows. In Sect.~\ref{Sect2}, the Hamiltonian model is introduced. Sections~\ref{Sect3} and \ref{Sect4} study the phase-space structures from numerical and analytical viewpoints, respectively. In Sect.~\ref{Sect5}, numerical maps are produced for flipping orbits, maximum eccentricity, and chaotic indicator. Finally, Sect.~\ref{Sect6} summarizes our main conclusions.

%%%%%%%%%%%%%%%%%%%%%%%%%%%%%%%%%%%%%%%%%%%%%%%%%%%%%%%%%%%%%%
\section{Hamiltonian Model}
\label{Sect2}

In this work, we consider a three-body dynamical model, wherein an inner test particle orbits a central star of mass $m_0$, perturbed by a distant companion of mass $m_{\text{p}}$. In the test-particle limit, the perturber moves around the central star on a fixed Keplerian orbit. For convenience, the system is formulated in an $m_0$-centered invariant plane coordinate frame, where the $x$- and $z$-axes are aligned with the perturber's eccentricity vector and angular momentum vector, respectively. In this frame, the orbital elements of the test particle (and the perturber) are denoted by the semi-major axis $a$ ($a_{\text{p}}$), eccentricity $e$ ($e_{\text{p}}$), inclination $i$ ($i_{\text{p}}$), longitude of ascending node $\Omega$ ($\Omega_{\text{p}}$), argument of pericenter $\omega$ ($\omega_{\text{p}}$), and mean anomaly $M$ ($M_{\text{p}}$).

In a hierarchical configuration, the semi-major axis ratio $\alpha = a / a_{\text{p}}$ is a small parameter, enabling the Hamiltonian to be expanded as a power series in $\alpha$ \citep{harrington1968,harrington1969}. Typically, truncating the Hamiltonian at order $\alpha^2$ yields the quadrupole approximation, whereas truncating at $\alpha^3$ leads to the octupole level. To study long-term dynamical evolution, short-period effects can be filtered out through double averaging of the Hamiltonian over orbital periods of both the test particle and the perturber \citep{ford2000,naoz2013}.

However, for systems with a weak hierarchy (specifically, when Equation (\ref{EqI1}) is violated), the double-averaging (DA) approximation may break down. To overcome this limitation, the (extended) Brown Hamiltonian, namely a nonlinear correction to the quadrupole-order Hamiltonian, can be employed \citep{brown1936stellar,cuk2004secular,breiter2015secular,luo2016,lei2018modified,will2021higher,tremaine2023,lei2025extensionsI,gao2025eccentric}. Under the DA framework, the test particle's semi-major axis is conserved, while its eccentricity and inclination undergo coupled evolution on timescales far exceeding orbital periods. When the eccentricity is excited to extreme values, the test particle makes close approaches to the central body, bringing short-range forces into play \citep{wu2003,fabrycky2007,liu2015,naoz2016,hamilton2021secular,huang2026}. As a representative example of short-range forces, GR precession is the primary focus of this work.

Incorporating both the Brown correction and GR precession, the double-averaged Hamiltonian up to octupole order can be written as \citep{ford2000,fabrycky2007,liu2015,luo2016,lithwick2011,li2014a,naoz2016,lei2022AJ,tremaine2023}
\begin{equation}\label{Eq2}
    \mathcal{H} = C_0\left(\mathcal{H}_{\text{quad}} + \epsilon_{\text{oct}} \mathcal{H}_{\text{oct}}+
    \epsilon_{\text{GR}} \mathcal{H}_{\text{GR}}+
    \epsilon_{\text{B}}\mathcal{H}_{\text{B}}\right),
\end{equation}
where the scaling coefficient $C_0$ is defined as
\begin{equation*}
    C_0 = \frac{3}{8} \frac{{\mathcal G} m_{\text{p}}}{a_{\text{p}}} \left( \frac{a}{a_{\text{p}}} \right)^2 \frac{1}{\left( 1 - e_{\text{p}}^2 \right)^{3/2}}.
\end{equation*}
Both $a$ and $a_{\text{p}}$ remain constant during secular evolution, thus $C_0$ is invariant within this DA framework. Consequently, the Hamiltonian can be conveniently normalized by $C_0$.

The explicit expressions for the quadrupole Hamiltonian $\mathcal{H}_{\text{quad}}$, octupole Hamiltonian $\mathcal{H}_{\text{oct}}$, GR perturbation $\mathcal{H}_{\text{GR}}$, and Brown correction $\mathcal{H}_{\text{B}}$ are provided in Appendix~\ref{SectA1}. The dimensionless coefficients $\epsilon_{\text{oct}}$, $\epsilon_{\text{GR}}$, and $\epsilon_{\text{B}}$ quantify the relative strength of the octupole, GR, and Brown terms with respect to the quadrupole potential, respectively, and are given by
\begin{equation*}
\begin{aligned}
    \epsilon_{\text{oct}} &= \frac{a}{a_{\text{p}}} \frac{e_{\text{p}}}{1 - e_{\text{p}}^2},\\
    \epsilon_{\text{GR}} &= \frac{8{\mathcal G} m_0^2 a_{\text{p}}^3 \left(1 - e_{\text{p}}^2\right)^{3/2}}{ m_{\text{p}} a^4 c^2},\\
   \epsilon_{\text{B}} &= \frac{m_{\text{p}}}{\sqrt{m_0 (m_0 + m_{\text{p}})}} \left(\frac{a}{a_{\text{p}}\left(1 - e_{\text{p}}^2\right)}\right)^{3/2} \left( 1 + \frac{2}{3} e_{\text{p}}^2 \right),
\end{aligned}
\end{equation*}
where ${\mathcal G}$ is the universal gravitational constant and $c$ is the speed of light.

For convenience, we introduce the normalized Delaunay variables by \citep{lithwick2011}
\begin{equation}\label{Eq3}
\begin{aligned}
    &g = \omega, \quad G = \sqrt{1 - e^2},\\
    &h = \Omega, \quad H = G \cos i,
\end{aligned}
\end{equation}
and express the normalized Hamiltonian as
\begin{equation}\label{Eq4}
\begin{aligned}
    \mathcal{H}(g, G, h, H) = & \mathcal{H}_{\text{quad}}(g, G, H) + \epsilon_{\text{oct}} \mathcal{H}_{\text{oct}}(g, G, h, H) \\
    & + \epsilon_{\text{GR}} \mathcal{H}_{\text{GR}}(G) + \epsilon_{\text{B}} \mathcal{H}_{\text{B}}(g, G, H).
\end{aligned}
\end{equation}
It is noticed that the Hamiltonian (\ref{Eq4}) can be collected as two components,
\begin{equation}\label{Eq4-1}
\mathcal{H}(g, G, h, H) =  \mathcal{H}_{\text{kernel}}(g, G, H) + \epsilon_{\text{oct}} \mathcal{H}_{\text{oct}}(g, G, h, H).
\end{equation}
where $\mathcal{H}_{\text{kernel}}$ is the kernel Hamiltonian, defined by
\begin{equation}\label{Eq4-2}
\mathcal{H}_{\text{kernel}} (g, G, H) = \mathcal{H}_{\text{quad}} + \epsilon_{\text{B}} \mathcal{H}_{\text{B}} + \epsilon_{\text{GR}} \mathcal{H}_{\text{GR}}.
\end{equation}
The kernel Hamiltonian determines an integrable dynamical model. From the viewpoint of perturbative treatment, the octupole-order Hamiltonian $\mathcal{H}_{\text{oct}}$ can be treated as a perturbation to the kernel Hamiltonian model. 

This formulation (\ref{Eq4}) governs a dynamical system with two degrees of freedom. Hamiltonian canonical equations then yield the governing equations of motion \citep{morbidelli2002modern}:
\begin{equation}\label{Eq5}
    \begin{aligned}
        &\frac{{\rm d}g}{{\rm d}t} = \frac{\partial \mathcal{H}}{\partial G}, \quad \frac{{\rm d}G}{{\rm d}t} = -\frac{\partial \mathcal{H}}{\partial g},\\
        &\frac{{\rm d}h}{{\rm d}t} = \frac{\partial \mathcal{H}}{\partial H}, \quad \frac{{\rm d}H}{{\rm d}t} = -\frac{\partial \mathcal{H}}{\partial h}.
    \end{aligned}
\end{equation}
For all numerical simulations performed in this work, we take a system with $\epsilon_{\text{oct}}=0.0188$, $\epsilon_{\text{GR}}=0.0084$, and $\epsilon_{\text{B}}=0.0103$ as a representative baseline example. We emphasize that the methodology presented here is general and readily applicable to different settings.

\section{Large-scale chaos driven by GR}
\label{Sect3}

In this section, we investigate the dynamical structures under the octupole-order Hamiltonian with GR precession and/or the Brown correction by constructing Poincar\'{e} surfaces of section, and identify the apsidal precession resonances under three different models by calculating the fundamental frequencies.

\subsection{Poincar\'{e} sections}
\label{Sect3-1}

Since the present Hamiltonian system possesses two degrees of freedom with only one known conserved quantity (the Hamiltonian itself), it is convenient to explore its phase-space structure by means of Poincar\'{e} surfaces of section \citep{lithwick2011,li2014a,lei2022AJ}. We define the Poincar\'{e} section by
\begin{equation*}
    g = 0,\quad \dot{g} > 0.
\end{equation*}
The trajectories are recorded each time they cross this surface of section. On the Poincar\'{e} section, smooth curves correspond to regular (quasi-periodic) orbits, the centers of libration islands represent stable periodic orbits, and scattered points signify chaotic regions \citep{lithwick2011,li2014a,lei2022AJ}.

\begin{figure*}%[ht!]
    \centering
    {\includegraphics[width=1.8\columnwidth]{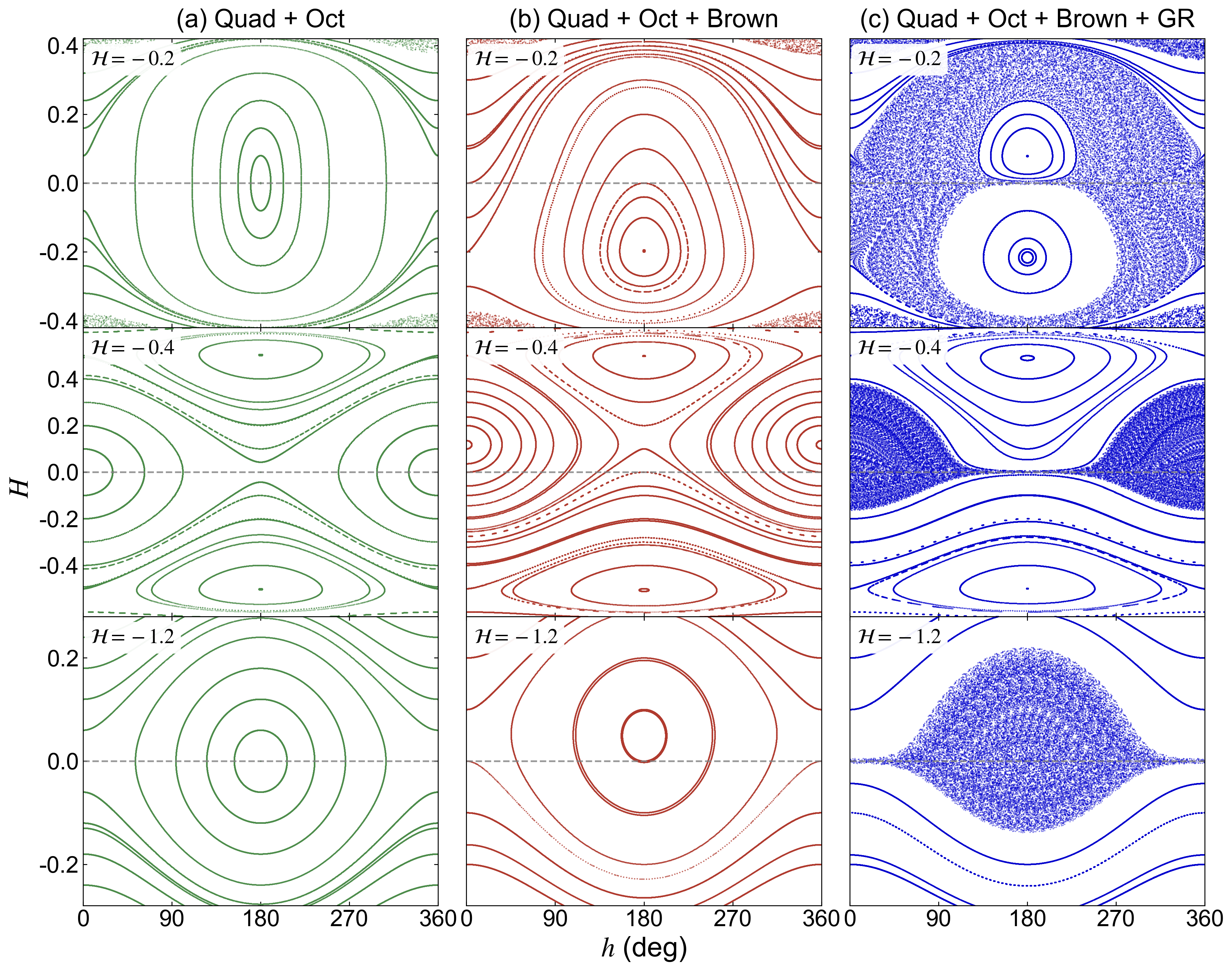}}
     \caption{Poincar\'{e} sections in the $(h, H)$ plane evaluated at different Hamiltonian levels: $\mathcal{H} = -0.2$ (\textit{top-row panels}), $\mathcal{H} = -0.4$ (\textit{middle-row panels}), and $\mathcal{H} = -1.2$ (\textit{bottom-row panels}). The left-, middle-, and right-column panels correspond to the `quadrupole + octupole', `quadrupole + octupole + Brown', and `quadrupole + octupole + Brown + GR' models, respectively. The gray dashed lines indicate $H = 0$ ($i = 90^\circ$).}
      \label{fig1}
\end{figure*}

To reveal how different physical perturbations affect the dynamical structure, Poincar\'{e} sections are constructed for various model configurations, including the `quadrupole + octupole', `quadrupole + octupole + Brown' and `quadrupole + octupole + Brown + GR' models. The results are shown in Fig.~\ref{fig1}.

For the `quadrupole + octupole' model (see the first column of Fig.~\ref{fig1}), the phase-space structure varies significantly with the level of Hamiltonian. At $\mathcal{H} = -0.2$, a single resonance island appears, centered at $(h = \pi, H = 0)$ (i.e., $i = 90^\circ$). As the Hamiltonian decreases to $\mathcal{H} = -0.4$, three distinct resonance islands emerge: one centered at $(h = 0, H = 0)$, and the other two symmetrically placed about $H = 0$ with centers at $(h = \pi, H \neq 0)$. When the Hamiltonian is further reduced to $\mathcal{H} = -1.2$, the dynamical structure reverts to a configuration similar to the first case, characterized by a single island at $(h = \pi, H = 0)$. In general, the trajectories residing within these islands are strictly regular (the chaotic layers are too thin to be resolved).

When the Brown correction is incorporated (Fig.~\ref{fig1}, middle column), the symmetry with respect to $H = 0$ is broken. Notably, however, both the number of resonance islands and the regularity of the trajectories within them remain unchanged.

By contrast, the inclusion of GR effects (Fig.~\ref{fig1}, right column) induces more pronounced changes in the phase-space structure. At $\mathcal{H} = -0.2$, the original libration island centered at $(h = \pi, H = 0)$ vanishes and is replaced by two distinct resonance islands at $(h = \pi, H \neq 0)$. Simultaneously, large-scale chaos develops in regions that were previously regular. At $\mathcal{H} = -0.4$, the central resonance island around $(h = 0, H = 0)$ dissolves into a chaotic domain, whereas the islands centered at $(h = \pi, H \neq 0)$ persist. Finally, at $\mathcal{H} = -1.2$, the original resonance island at $(h = \pi, H = 0)$ disappears, completely overtaken by a chaotic sea.

Poincar\'e sections reveal that the GR term significantly alters the phase-space structure, notably triggering widespread chaos. This naturally raises the central question of this work: What is the underlying mechanism driving the GR-enabled large-scale chaos?

\subsection{Fundamental structures under the kernel Hamiltonian models}
\label{Sect3-2}

The resonant argument $\sigma_1 = h + \mathrm{sign}(H)g$ is widely adopted to study secular resonance dynamics \citep{sidorenko2018,lei2022AA,lei2022AJ}. At the quadrupole order (the leading-order approximation), the nominal locations of secular resonances are determined by the behavior of $\sigma_1$ and its time derivative $\dot{\sigma}_1$ \citep{sidorenko2018,lei2022AJ}. Building upon this leading-order framework, we further investigate how the Brown correction and GR precession shift these resonance locations.

From the canonical equations of motion, the time derivative of $\sigma_1$ under the kernel Hamiltonian is given by
\begin{equation}\label{Eq6}
\dot{\sigma}_1 = \frac{\partial \mathcal{H}_{\text{kernel}}}{\partial H} + \mathrm{sign}(H) \frac{\partial \mathcal{H}_{\text{kernel}}}{\partial G}.
\end{equation}
Following the approach of \citet{lei2022AJ}, we average the precession rate $\dot{\sigma}_1$ over one period of a circulating ZLK cycle by
\begin{equation}\label{Eq7}
\langle \dot{\sigma}_1 \rangle = \frac{1}{T} \int_0^T \left[ \frac{\partial \mathcal{H}_{\text{kernel}}}{\partial H} + \mathrm{sign}(H) \frac{\partial \mathcal{H}_{\text{kernel}}}{\partial G} \right] {\rm d}t,
\end{equation}
where $T$ denotes the period of the ZLK cycle. Without loss of generality, we take ZLK cycles starting at $\omega_0 = 0$ with initial eccentricity $e_0$ and inclination $i_0$. The corresponding initial Delaunay variables are
\begin{equation*}
g_0 = 0,\quad G_0 = \sqrt{1 - e_0^2},\quad H_0 = G_0 \cos i_0.
\end{equation*}

For the `quadrupole' model, the kernel function $\mathcal{H}_{\text{kernel}}$ is taken as $\mathcal{H}_{\text{quad}}$. It is $\mathcal{H}_{\text{quad}} + \epsilon_{\text{B}}\mathcal{H}_{\text{B}}$ with the inclusion of Brown correction, and it is $\mathcal{H}_{\text{quad}} + \epsilon_{\text{B}}\mathcal{H}_{\text{B}} + \epsilon_{\text{GR}}\mathcal{H}_{\text{GR}}$ with additional inclusion of GR precession. Solving the condition of $\langle \dot{\sigma}_1 \rangle = 0$ yields the nominal locations of apsidal precession resonances for the respective models.

 \begin{figure*}%[ht!]
    \centering
    {\includegraphics[width=1.8\columnwidth]{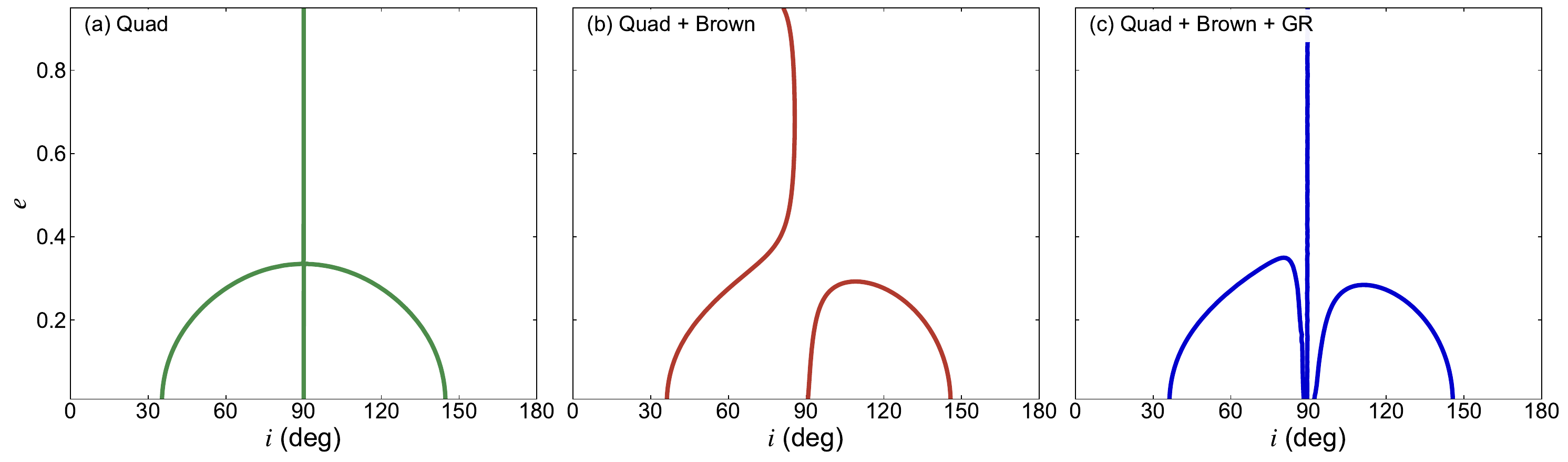}}
     \caption{Nominal location of apsidal precession resonances (defined by $\langle \dot{\sigma}_1 \rangle = 0$) in the initial inclination–eccentricity plane under three kernel Hamiltonian models: the `quadrupole' (\textit{left panel}), `quadrupole + Brown' (\textit{middle panel}), and `quadrupole + Brown + GR' (\textit{right panel}) models.}
      \label{fig2}
\end{figure*}  

Figure~\ref{fig2} presents the nominal locations of apsidal precession resonances in the initial inclination–eccentricity plane. In the quadrupole model (left panel), three distinct branches of resonance centers exist: one corresponding to polar orbits ($i_0 = 90^\circ$) across arbitrary eccentricities, and two symmetrically distributed about $i_0 = 90^\circ$ at low eccentricities, forming the asymmetric family. Upon incorporating the Brown term (middle panel), the branch at $i_0 = 90^\circ$ vanishes entirely, and the symmetry of the resonance distribution about $i_0 = 90^\circ$ is broken. With the further addition of GR effects (right panel), the $i_0 = 90^\circ$ branch reappears, accompanied by two asymmetric branches in the low-eccentricity regime alongside additional resonance structures near $i_0 = 90^\circ$. The predicted distribution of resonance centers aligns remarkably well with the Poincaré sections depicted in Fig.~\ref{fig1}.

\section{Analytical study on dynamical structures}
\label{Sect4}

In the previous section, we showed that GR precession significantly alters the nominal locations of apsidal precession resonances and enables large-scale chaos in phase space. The main goal of this section is to uncover the mechanism underlying this widespread chaos by applying analytical perturbation theory \citep{wisdom1985}.

\subsection{Perturbative treatment}
\label{Sect4-1}

To study the resonant structures associated with the critical argument $\sigma_1 = h + \mathrm{sign}(H)g$, we introduce a new set of canonical variables \citep{sidorenko2018,lei2022AJ,lei2022AA},
\begin{equation}\label{Eq8}
    \begin{aligned}
        &\sigma_1 = h + \mathrm{sign}(H)g, \quad \Sigma_1 = H,\\
        &\sigma_2 = g, \quad \Sigma_2 = G - |H|.
    \end{aligned}
\end{equation}
which is defined by the generating function,
\begin{equation}\label{Eq9}
    S = h \Sigma_1 + g (|\Sigma_1| + \Sigma_2).
\end{equation}
In the vicinity of apsidal precession resonance, this transformation effectively separates the resonant (slow) angle from the non-resonant (fast) angle.

In terms of these new variables, the Hamiltonian is written as
\begin{equation}\label{Eq10}
\mathcal{H} = \mathcal{H}_{\text{kernel}}(\sigma_2,\Sigma_1,\Sigma_2) + \epsilon_{\rm oct} \mathcal{H}_{\text{oct}}(\sigma_1,\sigma_2,\Sigma_1,\Sigma_2),
\end{equation}
where the kernel Hamiltonian $\mathcal{H}_{\text{kernel}}$, with inclusion of Brown correction and GR precession, is provided by Equation (\ref{Eq4-2}). 

The Hamiltonian (\ref{Eq10}) represents a system with distinct timescales, where $(\sigma_1, \Sigma_1)$ and $(\sigma_2, \Sigma_2)$ correspond to the slow and fast degrees of freedom, respectively. Consequently, when considering the evolution of the fast variables $(\sigma_2, \Sigma_2)$ over a single period, the slow variables $(\sigma_1, \Sigma_1)$ can be treated as approximately constant. This adiabatic approximation is widely adopted in diverse dynamical contexts \citep{wisdom1985,saillenfest2016,sidorenko2018,lei2022AJ,lei2022AA}. According to adiabatic invariant theory, the full system reduces to a one-degree-of-freedom (1-DOF) model on timescales much shorter than the period of the slow angle. For this integrable subsystem (where $\sigma_1$ and $\Sigma_1$ act as fixed parameters), action-angle variables can be introduced by \citep{morbidelli2002modern}
\begin{equation}\label{Eq11}
    \Sigma_2^* = \frac{1}{2\pi} \int_0^{2\pi} \Sigma_2 \, {\rm d}\sigma_2, \quad \sigma_2^* = \frac{2\pi}{T} t,
\end{equation}
where $\Sigma_2^*$ denotes the path integral of the solution curve $\Sigma_2(\sigma_2)$ over one complete period, divided by $2\pi$, and $\sigma_2^*$ is a linear function of time (with $T$ the period of $\sigma_2$).

By introducing the action variable, the Hamiltonian given by Equation~(\ref{Eq10}) can be rewritten as
\begin{equation}\label{Eq12}
    \mathcal{H}(\sigma_1, \Sigma_1, \Sigma_2^*) = \mathcal{H}_{\text{kernel}}(\Sigma_1, \Sigma_2^*) + \epsilon_{\rm oct} \mathcal{H}_{\text{oct}}(\sigma_1, \Sigma_1, \Sigma_2^*),
\end{equation}
which governs the long-term evolution of the slow variables. In this formulation, the fast angle $\sigma_2^*$ is cyclic (absent from the Hamiltonian), rendering its conjugate momentum $\Sigma_2^*$ an integral of motion on secular timescales. Consequently, the reduced system possesses two independent integrals of motion, namely the Hamiltonian $\mathcal{H}$ itself and the action $\Sigma_2^*$, and is therefore integrable. For a fixed value of the Hamiltonian, phase portraits can be generated by plotting level curves of the adiabatic invariant $\Sigma_2^*$ in the $(\sigma_1, \Sigma_1)$ phase space.

\begin{figure*}%[ht!]
    \centering
    {\includegraphics [width=1.8\columnwidth]{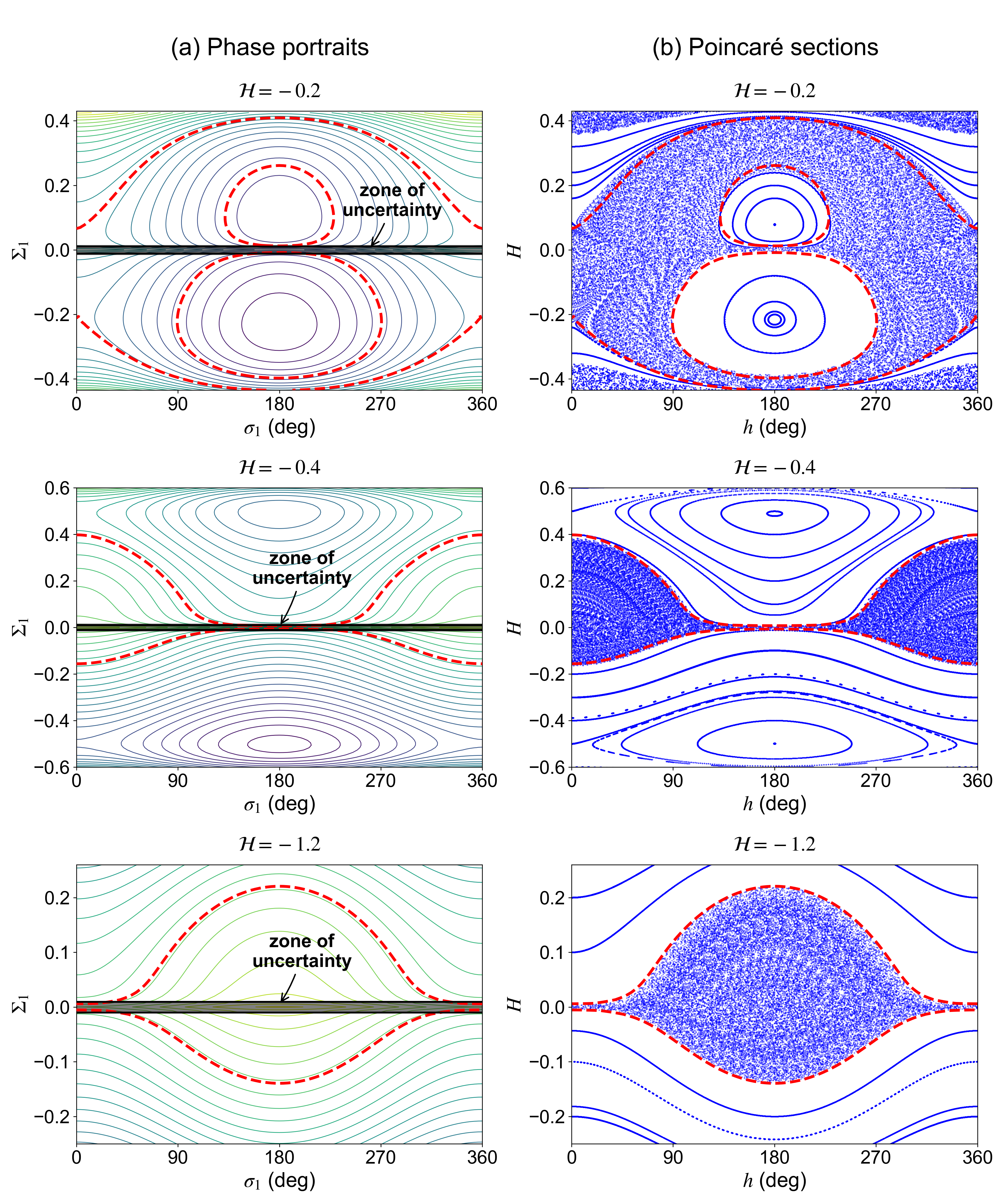}}
     \caption{Phase portraits (\textit{left-column panels}) and the corresponding Poincaré sections (\textit{right-column panels}) at three Hamiltonian levels: $\mathcal{H}=-0.2$ (\textit{top-row panels}), $\mathcal{H}=-0.4$ (\textit{middle-row panels}), and $\mathcal{H}=-1.2$ (\textit{bottom-row panels}). The gray-shaded region in the left-column panels highlights the zone of uncertainty, bounded by the contour of $\log_{10}(1 - e_{\max}) = -4.5$, and trajectories traversing this zone are identified as chaotic. The red dashed lines mark the analytical boundaries of chaos derived from the phase-portrait analysis.}
      \label{fig3}
\end{figure*}  

Figure~\ref{fig3} presents the phase portraits (left panels) and the corresponding Poincar\'e surfaces of section (right panels) for three different levels of Hamiltonian at $\mathcal{H} = -0.2, -0.4, -1.2$. Overall, the dynamical structures shown in the phase portraits closely mirror those observed in the Poincar\'{e} sections. 

In regions where the Poincaré sections exhibit chaotic motion, the corresponding phase-portrait trajectories invariably intersect zones of uncertainty\footnote{The same terminology as that of \citet{wisdom1985} is used in the present study to understand the chaotic mechanism.}, marked by gray-shaded regions (see the left-column panels of Fig.~\ref{fig3}). Within this zone, it is observed that the level curves of the adiabatic invariant $\Sigma_2^*$ become extremely dense.

It should be noted that, inside the uncertainty zone, the adiabatic invariant condition breaks down \citep{wisdom1985}. Notably, the uncertainty zone coincides with the regime where the double-averaging (DA) approximation fails and non-adiabatic jumps occur \citep{hamilton2024,klein2026}; this correspondence is demonstrated explicitly in Appendix~\ref{SectA5}. In this regime, orbits can transition between different DA trajectories on the outer orbital timescale, leading to chaos on secular timescales.

According to \citet{wisdom1985}, we demarcate this neighborhood with a gray-shaded region in the phase portraits. 
Quantitatively, its boundary is determined by the condition $\log_{10}(1 - e_{\max})=-4.5$, where $e_{\max}$ is the maximum eccentricity achieved over one ZLK cycle \footnote{For simplicity, the maximum eccentricity is evaluated from the kernel Hamiltonian. During one ZLK cycle, $e$ reaches its maximum $e_{\max}$ when the argument of pericenter is equal to $\omega_{e_{\max}} = \pi/2$.}. Away from this zone, the action variable serves as a good approximate integral, rendering the motion fully predictable: trajectories follow the contours of $\Sigma_2^*$ until they enter the uncertainty zone. Within this region, the dynamics become complex. Eventually, trajectories re-emerge along adjacent level curves with altered values of $\Sigma_2^*$ (see Appendix~\ref{SectA3} for a representative chaotic trajectory), and this process repeats. Each time a trajectory enters the uncertainty zone, it is forced toward the separatrix, where motion is inherently chaotic \citep{wisdom1985,lithwick2011}. These repeated excursions into the uncertainty zone are responsible for the observed large-scale chaos. This mechanism of producing chaos is in agreement with that described in \citet{wisdom1985}.

Consequently, trajectories in the phase portraits that cross this uncertainty zone are identified as chaotic orbits. The boundaries of these chaotic domains are determined by the outermost level curves entering the uncertainty zone, marked by red dashed curves. Overlaying these analytically derived chaotic boundaries onto the associated Poincar\'{e} sections reveals that they provide precise boundaries of chaotic sea (see the right-column panels of Fig.~\ref{fig3}).

\subsection{Analytical boundaries of chaos}
\label{Sect4-2}

\begin{figure*}%[ht!]
    \centering
    {\includegraphics [width=1.8\columnwidth]{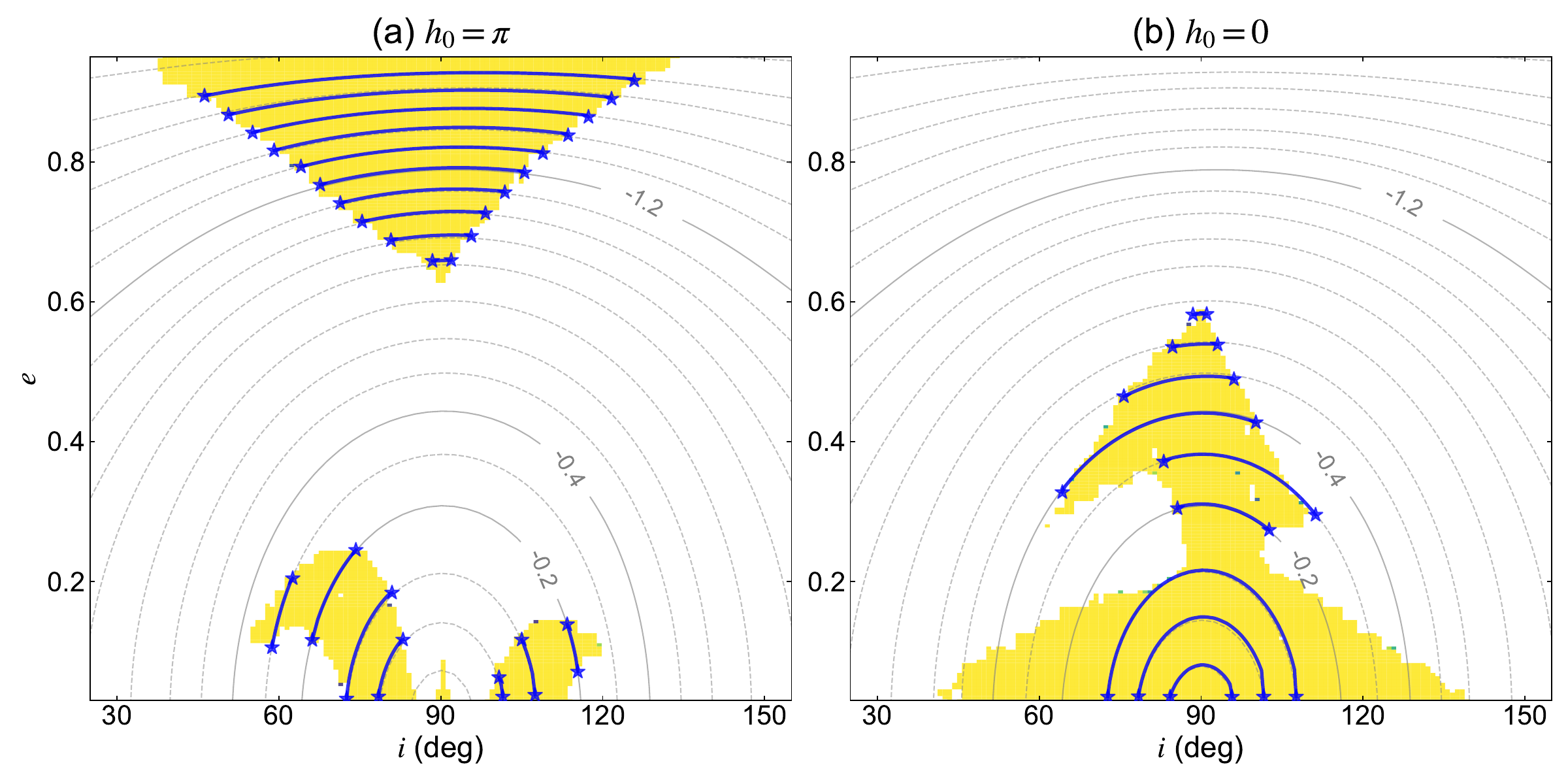}}
     \caption{Analytical chaotic boundaries (blue stars and lines) overlaid on the numerically determined chaotic regions (yellow areas). The yellow regimes are bounded by the level curves of $\log_{10} \mathrm{FLI} = 1.5$, where the chaotic indicator FLI is to be discussed in Sect.~\ref{Sect5}. Initial conditions are set at $g_0 = 0, h_0 = \pi$ in the \textit{left panel} and $g_0 = 0, h_0 = 0$ in the \textit{right panel}. Background contours depict the Hamiltonian level curves, with the three solid curves highlighting the energy levels $\mathcal{H}=-0.2, -0.4,$ and $-1.2$ examined in Figs.~\ref{fig1} and \ref{fig3}.}
      \label{fig4}
\end{figure*} 

As revealed by phase portraits, the center of the libration island around which chaos develops is located at either $0$ or $\pi$, depending on the level of Hamiltonian. Based on these resonance centers, the chaotic regions can be classified into two distinct types: one where the critical argument $\sigma_1$ librates around $\pi$ (corresponding to $h_0=\pi$), and another where it librates around $0$ (corresponding to $h_0=0$). These two cases are analyzed separately. Following the methodology introduced in the previous section, the chaotic boundaries at various Hamiltonian levels are extracted from the phase portraits by identifying the outermost level curves that enter the uncertainty zone. The resulting chaotic width, quantified by $\Delta \Sigma_1$, directly provides the lower and upper boundaries of the chaotic domain in the initial eccentricity–inclination space.

By mapping these chaotic boundaries onto the $(e_0, i_0)$ plane, we obtain the analytical boundaries of chaos, represented by the blue stars in Fig.~\ref{fig4}. In this figure, the chaotic widths are evaluated along the level curves of Hamiltonian, which are shown as the background. For convenience of comparison, the chaotic regions extracted from numerical maps are overlaid as yellow shaded areas in Fig.~\ref{fig4} (the detailed methodology for identifying these regions using a chaotic indicator will be discussed in the next section). The initial conditions for these simulations are assumed at $(g_0 = 0, h_0 = \pi)$ for the left panel and $(g_0 = 0, h_0 = 0)$ for the right panel. It is observed that the analytical predictions agree well with the numerical simulations, confirming that the large-scale chaos originates from trajectories crossing the uncertainty zone. In particular, chaotic orbits centered at $\sigma_1 = \pi$ are predominantly distributed in the low- and high-eccentricity regimes, whereas those centered at $\sigma_1 = 0$ mainly occupy the low- and moderate-eccentricity regimes.

However, minor discrepancies exist between the analytical and numerical results in the low-eccentricity regime for the case of $h_0 = 0$ (see the bottom-left and right corners). It should be noted that this chaos does not arise from GR effects; rather, it originates from those octupole-order resonances in the low-eccentricity region. This feature corresponds to the chaotic regions observed in the first row of Fig.~\ref{fig1} at $h = 0$ and high values of $|H|$ ($\sim 0.4$), which is outside the scope of the present study.

%%%%%%%%%%%%%%%%%%%%%%%%%%%%%%%%%%%%%%%%%%%%%%%%%%%%%%%%%%%%%%

\section{Flipping orbits, maximum eccentricity, and chaotic indicator}
\label{Sect5}

The preceding phase-space analysis has revealed the dynamical mechanism by which the GR term induces chaos, and has provided the analytical boundaries of chaos by analyzing phase portraits. According to the definition of $\Sigma_1 = H = G \cos i$, the line $\Sigma_1 = 0$ corresponds to $i = 90^\circ$; thus, trajectories inside the uncertainty zone near $\Sigma_1 = 0$ may flip between prograde and retrograde orbits. Furthermore, as demonstrated in Fig.~\ref{fig3}, all chaotic regions centered at $\Sigma_1 = 0$ are capable of orbital flips.

\begin{figure*}%[h!]
    \centering
    {\includegraphics[width=1.8\columnwidth]{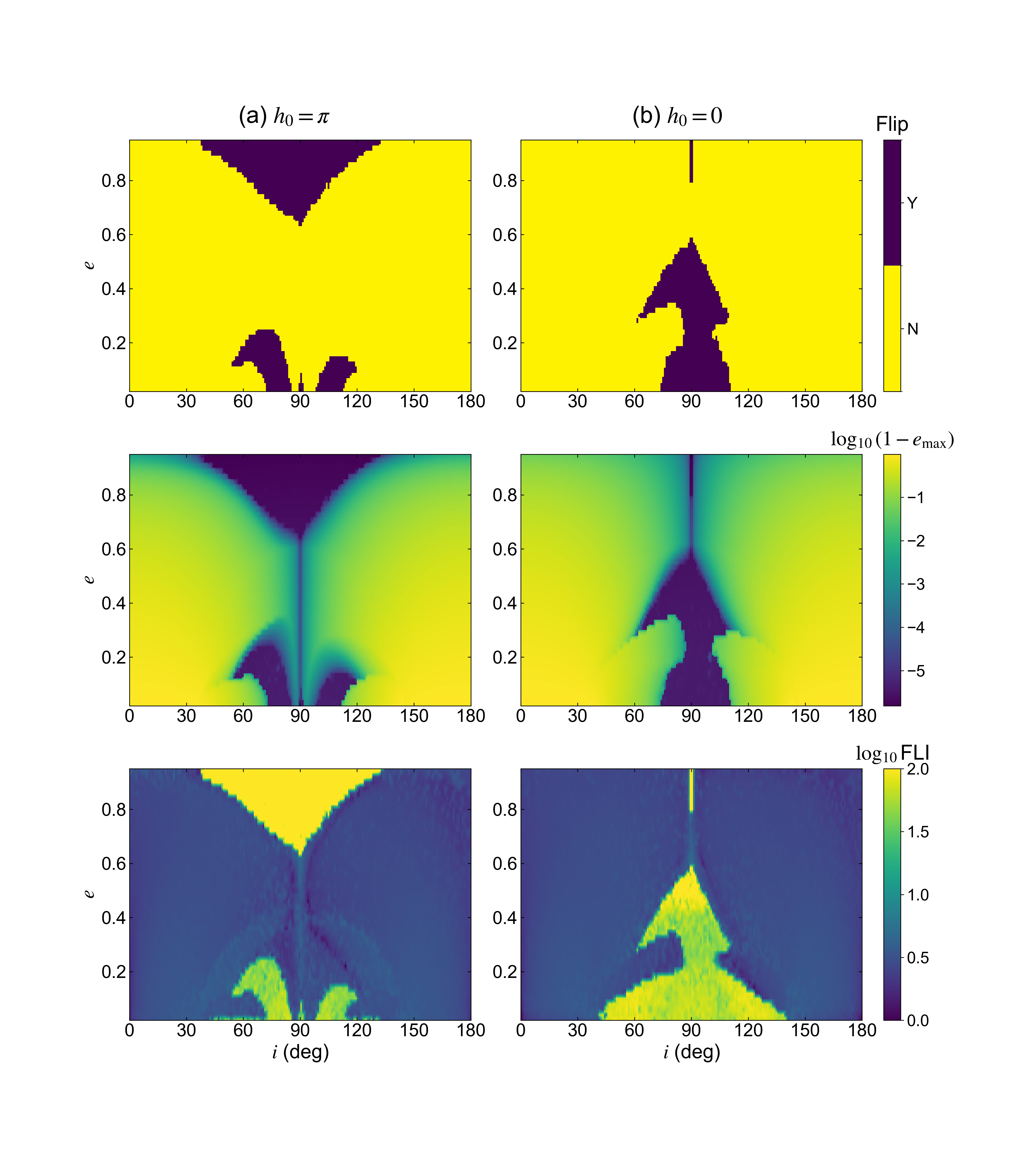}}
     \caption{Numerical maps showing the distribution of flipping orbits (\textit{top panels}), maximum eccentricity attained (\textit{middle panels}), and the chaos indicator FLI (\textit{bottom panels}) in the initial eccentricity–inclination plane. The left and right columns correspond to initial angle configurations of $(g_0 = 0, h_0 = \pi)$ and $(g_0 = 0, h_0 = 0)$, respectively.}
      \label{fig5}
\end{figure*}

To corroborate these findings, the equations of motion (\ref{Eq5}) are numerically integrated across the initial parameter space $(e_0, i_0)$ for specified initial angles $(g_0, h_0)$. In particular, we examine three dynamical features: the distribution of flipping orbits, the maximum eccentricity attained, and the map of chaotic indicator. The numerical results are summarized in Fig.~\ref{fig5}, where the left and right columns correspond to the initial conditions $(g_0 = 0, h_0 = \pi)$ and $(g_0 = 0, h_0 = 0)$, respectively.

As the chaotic indicator, we adopt the Fast Lyapunov Indicator (FLI), defined as \citep{skokos2009lyapunov}
\begin{equation}\label{Eq13}
    \mathrm{FLI}(k\tau) = \sup_k \sum_{i=1}^k \ln \frac{D_i}{D_0},
\end{equation}
where $\tau$ denotes the integration step size, $D_0$ is the norm of the initial deviation vector, and $D_i$ denotes the norm of the deviation vector after the $i$-th step. Regular orbits yield slowly growing FLI values, whereas chaotic orbits exhibit exponential growth, thereby providing a clear distinction between the two dynamical regimes.

Figure~\ref{fig5} shows, from top to bottom, the distribution of flipping orbits, the maximum eccentricity expressed as $\log_{10}(1 - e_{\text{max}})$, and the corresponding $\log_{10}\mathrm{FLI}$ map. In the top-row panel, yellow and purple regions represent non-flipping and flipping orbits, respectively. The flip regions exhibit a strong morphological resemblance to the chaotic regions centered at $\Sigma_1 = 0$ in Fig.~\ref{fig5}, confirming that, for the GR-dominated chaotic regime explored in this study, all flip regions are chaotic and both originate from trajectories crossing the uncertainty zone around $\Sigma_1 = 0$. Furthermore, regions experiencing extreme eccentricity excitation show a strong spatial overlap with both the flip and chaotic domains, indicating that orbital flips and chaotic motion are intrinsically accompanied by substantial eccentricity growth. We have also performed additional simulations with increasing $\epsilon_{\rm GR}$ to examine its influence on the chaotic regions. Please refer to Appendix~\ref{SectA4} for the results.

\begin{figure*}%[h!]
    \centering
{\includegraphics[width=1.7\columnwidth]{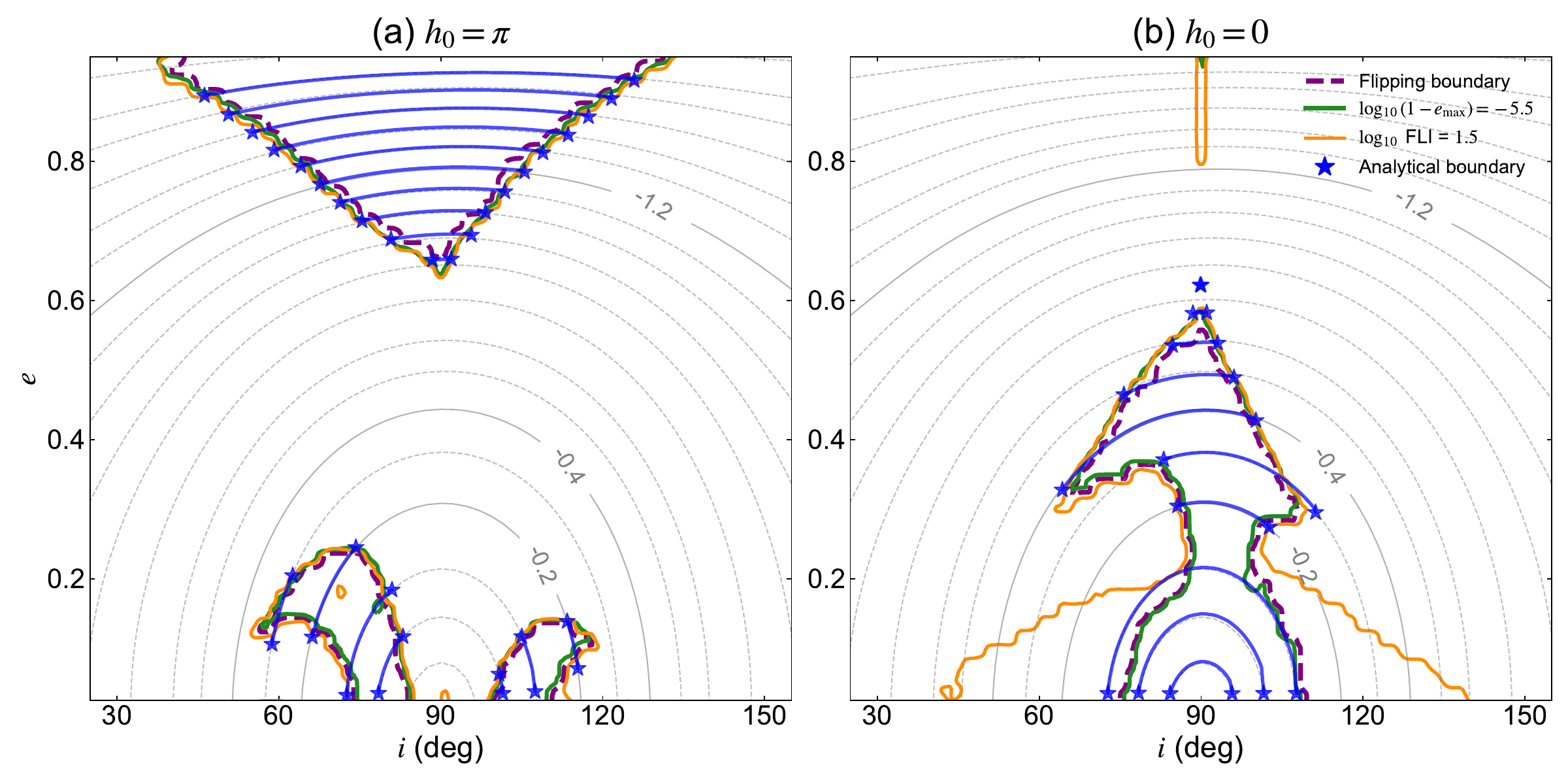}}
     \caption{Direct comparison among the numerical flipping boundary, the critical eccentricity contour ($\log_{10}(1-e_{\text{max}}) = -5.5$), the FLI chaotic boundary ($\log_{10} \mathrm{FLI} = 1.5$), and the analytical chaotic boundary derived from phase portraits. The left and right panels correspond to $h_0 = \pi$ and $h_0 = 0$, respectively (both with $g_0 = 0$). In both panels, Hamiltonian level curves are shown as the background, where the three energy levels $\mathcal{H}=-0.2, -0.4,$ and $-1.2$ examined in Figures~\ref{fig1} and \ref{fig3} are explicitly highlighted by gray solid lines. The threshold values for $\log_{10}(1-e_{\text{max}}) = -5.5$ and $\log_{10} \mathrm{FLI} = 1.5$ are established in Appendices~\ref{SectA2} and \ref{SectA3}.}
      \label{fig6}
\end{figure*}

To further verify the consistency between numerical results and analytical predictions, Fig.~\ref{fig6} directly overlays the numerically determined flip boundary, chaotic boundary, and high-eccentricity contour against the analytically predicted chaotic boundary. The purple dashed line indicates the flip boundary, and the green solid line, corresponding to $\log_{10}(1 - e_{\text{max}}) = -5.5$, represents the theoretical critical eccentricity threshold evaluated for orbital flips (see Appendix~\ref{SectA2} for details). The yellow solid line marks the chaotic boundary identified via the FLI criterion (detailed in Appendix~\ref{SectA3}), and the blue asterisks denote the analytical chaotic boundary derived from the phase-portrait analysis.

As shown in Fig.~\ref{fig6}, the excellent agreement among these four independently derived boundaries validates the perturbation theory developed here and connects these core dynamical features (including flipping orbits, maximum eccentricity, chaotic indicator, and analytical structure), thereby providing a solid framework for uncovering the mechanism behind GR-enabled large-scale chaos.

%%%%%%%%%%%%%%%%%%%%%%%%%%%%%%%%%%%%%%%%%%%%%%%%%%%%%%%%%%%%%%
\section{Conclusions}
\label{Sect6}

In this work, we systematically investigate the chaotic behavior of test particles by means of Poincar\'{e} surfaces of section, perturbation theory, and chaotic indicator, within the framework of the octupole-level secular approximation incorporating GR precession and the Brown correction.

Poincar\'{e} surfaces of section reveal that, in the purely Newtonian gravitational model, the phase space exhibits well-ordered resonant island structures. The inclusion of the Brown term breaks the dynamical symmetry of these structures. When GR precession is further incorporated, a widespread chaotic sea emerges, establishing the GR term as the key enabler of the system's chaotic dynamics. 

To investigate the chaotic properties of the phase space, we introduce a canonical transformation that separates the Hamiltonian into fast- and slow-varying dynamical components. For such a multi-timescale system, a perturbation framework based on the adiabatic invariant approximation is highly effective. The adiabatic invariant is defined as the phase-space area enclosed by the trajectory of the fast DOF. Consequently, with two independent integrals of motion—the total Hamiltonian and the adiabatic invariant—the reduced system becomes integrable. The corresponding phase portraits are constructed by plotting the level curves of adiabatic invariant at a specified Hamiltonian energy level. Our analysis of these phase portraits reveals that the inclusion of GR precession gives rise to a critical uncertainty zone near the polar orbit. Trajectories traversing this zone are identified as chaotic, and the resulting analytical chaotic boundaries align remarkably well with those mapped via Poincar\'e surfaces of section. 

Through direct numerical integration, we produce numerical maps of flipping orbits, the maximum eccentricity attained, and chaotic indicator in the initial eccentricity–inclination plane. Our primary results are presented in Fig.~\ref{fig6}, where the numerically computed flip boundary, the critical maximum eccentricity contour associated with orbital flips, and the chaotic boundary all show excellent agreement with our analytical predictions. 

In restricted hierarchical three-body systems with short-range forces, flipping orbits driven to extreme eccentricities are inherently chaotic. The underlying mechanism governing this large-scale chaos stems from the periodic traversal of trajectories through the phase-space uncertainty zone over secular timescales. This process is dynamically analogous to the mechanism responsible for chaotic dynamics inducing the 3/1 Kirkwood gap in the Solar System \citep{wisdom1985}.

%%%%%%%%%%%%%%%%%%%%%%%%%%%%%%%%%%%%%%%%%%%%%%%%%%%%%%%%%%%%%%
\begin{acknowledgements}
We thank the anonymous reviewer for his/her insightful comments, which have substantially improved the quality of this work. This work is financially supported by the National Natural Science Foundation of China (Nos. 12573063, 12233003 and 12073011) and the China Manned Space Program with grant no. CMS-CSST-2025-A16.
\end{acknowledgements}

\bibliographystyle{aa}
\bibliography{mybib}

\begin{appendix}
	
\nolinenumbers

\section{Explicit expressions of the Hamiltonian}
\label{SectA1}

The explicit expressions of the double-averaged Hamiltonian with Brown correction and GR precession can be found in previous references. In particular, the quadrupole-order term is given by \citep{kozai1962,lidov1962}
\begin{equation}
    \mathcal{H}_{\text{quad}} = - \left(1 + \frac{3}{2}e^2\right) \theta^2 + \frac{1}{2} e^2\left[1 - 5\left(1 - \theta^2\right)\cos{2\omega}\right],
\end{equation}
and the octupole-order term is \citep{lithwick2011,naoz2016,klein2024hierarchicalI}
\begin{equation}
\begin{aligned}
    \mathcal{H}_{\text{oct}} =&- \frac{5}{16} e \left( 1 + \frac{3}{4} e^2 \right) \\
    &\times \left[ \left( 1 - 11 \theta - 5 \theta^2 + 15 \theta^3 \right) \cos (\omega - \Omega) \right. \\
    &+ \left. \left( 1 + 11 \theta - 5 \theta^2 - 15 \theta^3 \right) \cos (\omega + \Omega) \right] \\
    &- \frac{175}{64} e^3 \left[ \left( 1 - \theta - \theta^2 + \theta^3 \right) \cos (3\omega - \Omega) \right. \\
    &+ \left. \left( 1 + \theta - \theta^2 - \theta^3 \right) \cos (3\omega + \Omega) \right],
\end{aligned}
\end{equation}
where $\theta = \cos i$. The potential due to GR effect is \citep{wu2003,liu2015}
\begin{equation}
    \mathcal{H}_{\text{GR}} = -\frac{1}{\eta},
\end{equation}
and the Brown term is \citep{luo2016,lei2018modified,tremaine2023,klein2024hierarchical,gao2025eccentric,lei2025extensionsI,lei2025extensionsII,lei2026extensions} 
\begin{equation}
    \mathcal{H}_{\text{B}} = -\frac{3}{16} \eta \, \theta \left[ 2(1-\theta^2) + e^2 \left( 33 + 17 \theta^2 + 15(1-\theta^2) \cos 2\omega \right) \right],
\end{equation}
where $\eta = \sqrt{1-e^2}$.

\section{Determination of the critical maximum eccentricity for orbit flips}
\label{SectA2}

In Fig.~\ref{fig6}, we adopt $\log_{10}(1 - e_{\text{max}}) = -5.5$ as the critical maximum eccentricity for achieving orbital flips. Here we provide a detailed consideration on how this criterion is determined and derived.

At the flipping instant ($i = 90^\circ$), the eccentricity reaches its maximum. In addition, both the octupole-order term and Brown correction term vanish at $i = 90^\circ$. Thus the maximum eccentricity $e_{\text{max}}$ from the `quadrupole + octupole + GR +Brown' model is consistent with the limiting eccentricity $e_{\text{lim}}$ derived from the `quadrupole + GR' model \citep{liu2015,huang2026}. This is the reason that we could take $e_{\text{lim}}$ as a reasonable approximation for the maximum eccentricity threshold required for orbital flips.

From the conservation of the `quadrupole + Brown' Hamiltonian, we have \citep{huang2026}
\begin{equation}
\begin{aligned}
    &\mathcal{H}_{\text{quad}}(\omega_0, e_0, i_0=90^\circ) + \mathcal{H}_{\text{GR}}(e_0) = \\
    &\mathcal{H}_{\text{quad}}(\omega_{\text{elim}}, e_{\text{lim}}, i_{\text{elim}}=90^\circ) + \mathcal{H}_{\text{GR}}(e_{\text{lim}}).
\end{aligned}
    \label{eqa1}
\end{equation}
For a given initial eccentricity $e_0$, the maximum eccentricity is reached when $\omega_0 = 0$ or $\pi$, at which point the argument of pericenter becomes $\omega_{\text{elim}} = \pm \pi/2$. Substituting $\omega_0 = 0$ and $\omega_{\text{elim}} = \pm \pi/2$ into Equation (\ref{eqa1}) then yields the analytical expression for the critical maximum eccentricity \citep{liu2015,huang2026}:
\begin{equation}\label{Eq_B2}
    2e_0^2 + \frac{8\epsilon_{\text{GR}}}{3\sqrt{1 - e_0^2}} - \frac{8\epsilon_{\text{GR}}}{3\sqrt{1 - e_{\text{lim}}^2}} + 3e_{\text{lim}}^2 = 0.
\end{equation}

\begin{figure}%[ht!]
   \centering
   \includegraphics[width=0.8\columnwidth]{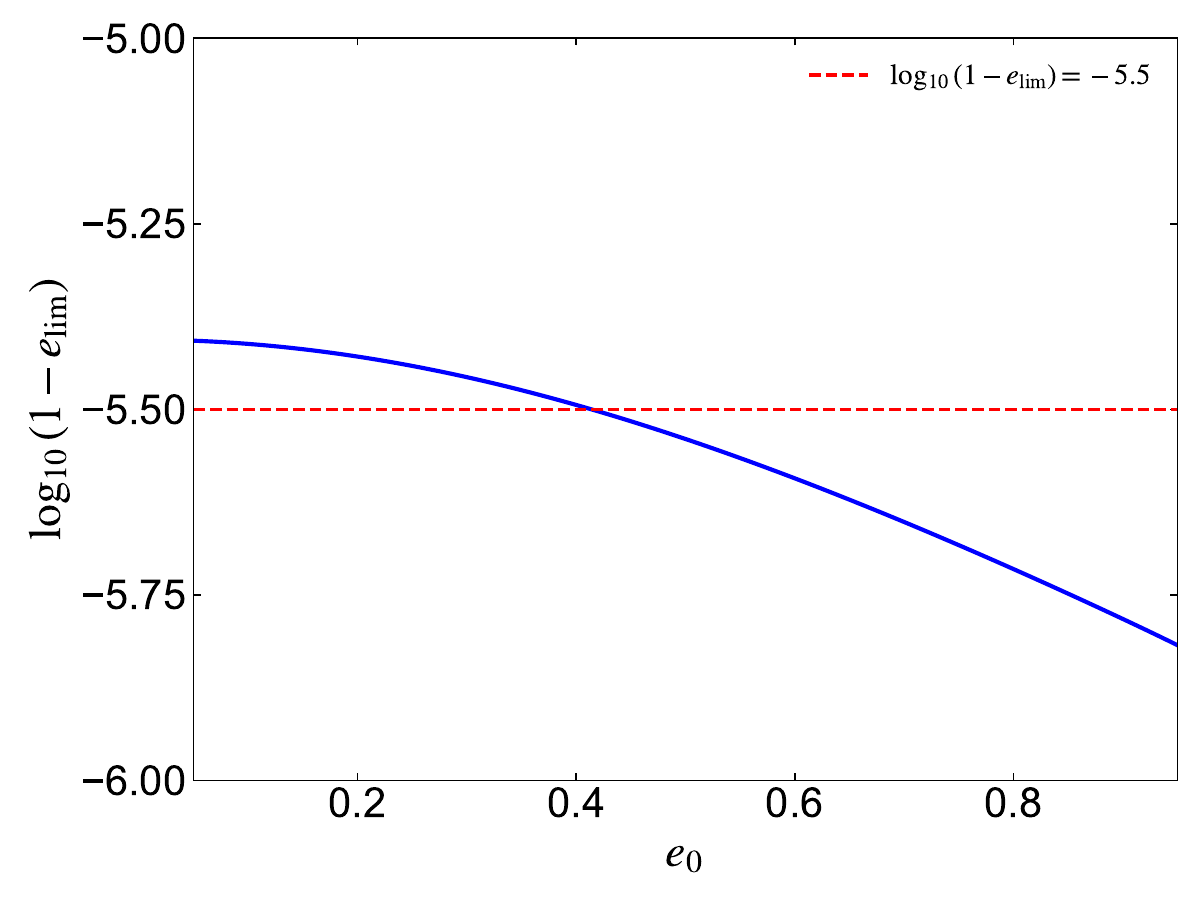}
      \caption{Limiting eccentricity $e_{\text{lim}}$ as a function of initial eccentricity $e_0$, obtained by solving Equation~(\ref{Eq_B2}). The mean value, $\log_{10}(1 - e_{\text{lim}}) = -5.5$, is indicated by the horizontal dashed line and is adopted as the criterion for the critical eccentricity reached by flipping orbits in Fig.~\ref{fig6}.}
         \label{figA1}
\end{figure}

Figure~\ref{figA1} shows the relationship between $\log_{10}(1 - e_{\text{lim}})$ and $e_0$ for the given $\epsilon_{\text{GR}}$. It can be observed that as $e_0$ increases, $e_{\text{lim}}$ approaches unity more closely, and the corresponding $\log_{10}(1 - e_{\text{lim}})$ decreases from $-5$ to $-6$. In this work, we adopt an approximate mean $\log_{10}(1 - e_{\text{lim}}) = -5.5$ as the criterion for the critical eccentricity reached by flipping orbits, which is used in Fig.~\ref{fig6}.

\section{Numerical validation of chaotic regions}
\label{SectA3}

To verify the reliability of the chaotic regions identified by the FLI method in the main text, here we apply the classical algorithm proposed by \cite{benettin1980} to compute the maximal Lyapunov exponent, and also specifies the FLI threshold adopted in Fig.~\ref{fig6} to distinguish chaotic from regular orbits.

By simultaneously integrating the original system and the tangent map with periodic Gram–Schmidt orthonormalization, we stably extract the full set of Lyapunov characteristic exponents (LCE). Here we focus solely on the maximal Lyapunov exponent as a chaos indicator: a positive value signals that the corresponding orbit is chaotic. 

\begin{figure*}%[h!]
   \centering
   \includegraphics[width=1.6\columnwidth]{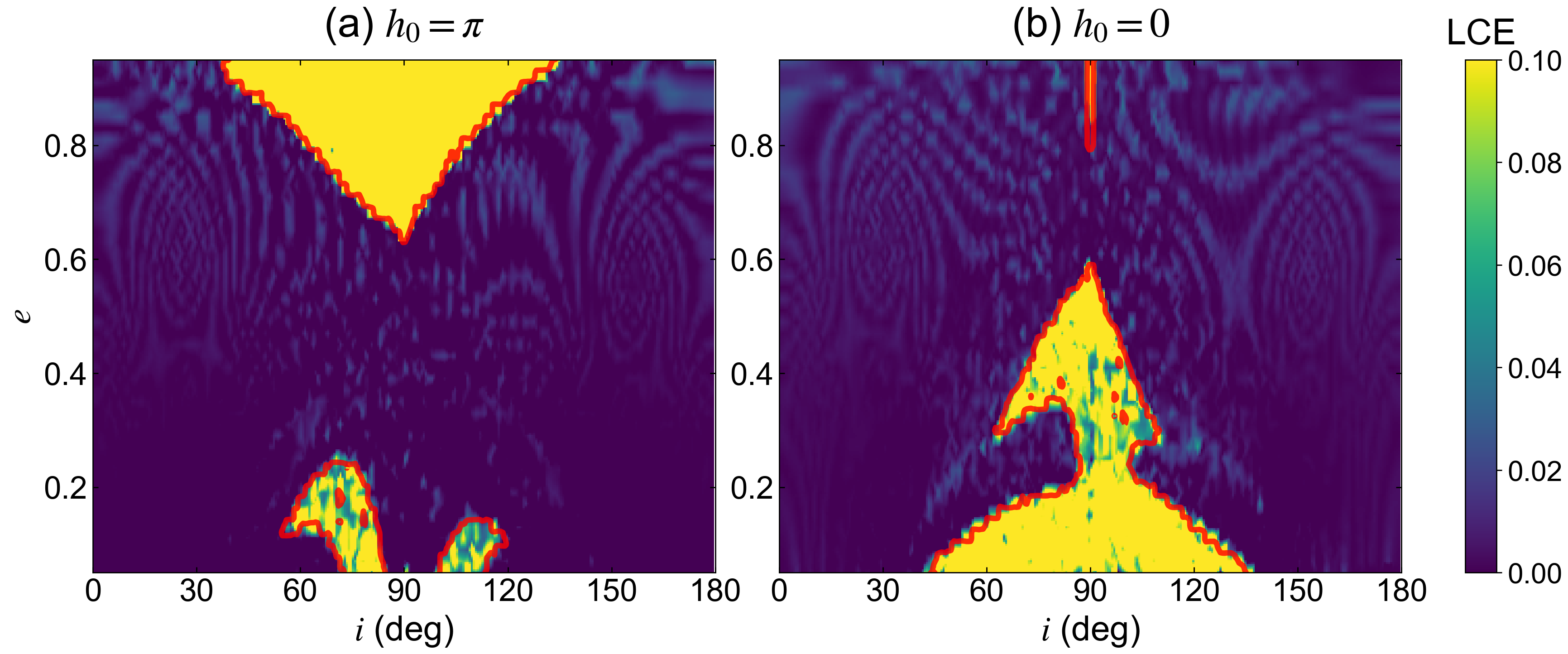}
      \caption{Distribution of the maximal Lyapunov exponent (LCE) in the initial eccentricity–inclination plane. The red curves denote the chaotic boundaries derived from the FLI (consistent with those in Fig.~\ref{fig6}). The excellent agreement between the FLI-based and LCE-based boundaries validates the consistency of the chaos indicators employed in this study.}
         \label{figC1}
\end{figure*}

Figure~\ref{figC1} presents the distribution of the maximal Lyapunov exponent in the $(e,i)$ plane, which closely matches the chaotic regions delineated by the FLI method (red dashed line), confirming the effectiveness of the FLI method for identifying chaotic regions in the main text.

\begin{figure*}%[ht!]
   \centering
   \includegraphics[width=1.8\columnwidth]{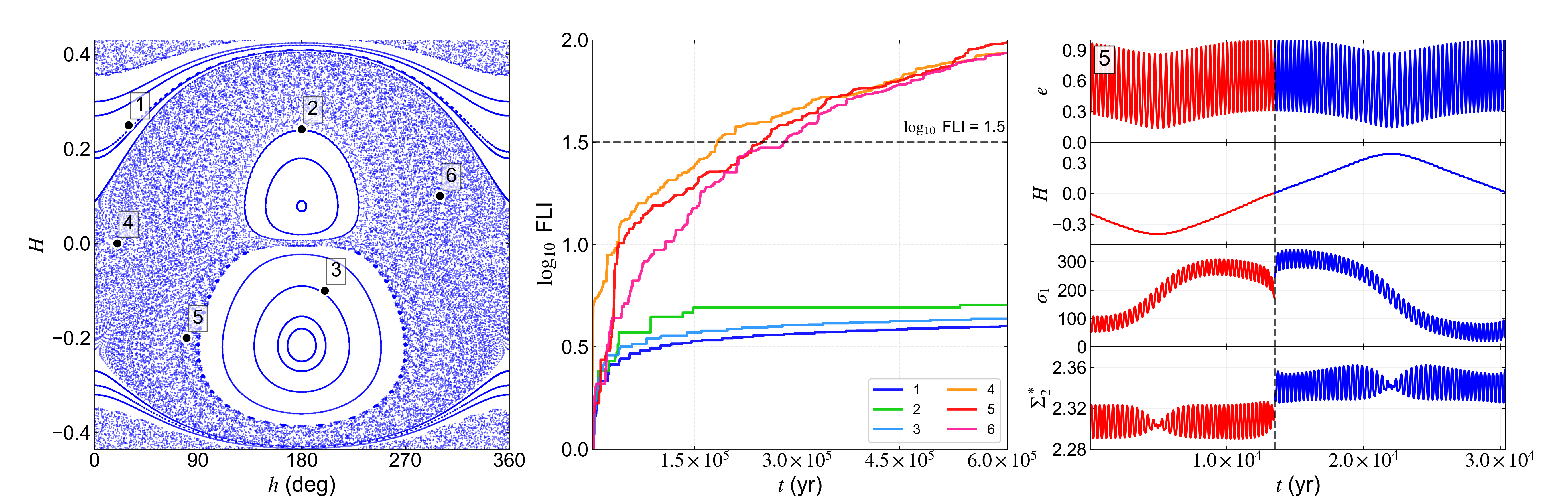}
      \caption{Poincar\'e surface of section at $\mathcal{H} = -1.2$ (left panel), time histories of FLI for representative orbits (middle panel), and time histories of $e$, $H$, $\sigma_1$, and $\Sigma_2^*$ for the chaotic trajectory labeled 5 (right panel). Trajectories labeled 1--3 reside within the regular regime, whereas those labeled 4--6 lie inside the chaotic region. The horizontal black dashed line in the middle panel marks the threshold $\log_{10} \mathrm{FLI} = 1.5$, which is adopted as the criterion for chaotic motion in Fig.~6. In the right panel, the vertical black dashed lines indicate the instants when the trajectory enters the uncertainty zone, during which $\Sigma_2^*$ undergoes significant changes.}
         \label{figC2}
\end{figure*}

To determine the FLI threshold for identifying chaos, we integrate representative trajectories originating from the Poincaré section and trace the temporal evolution of their FLIs. For trajectories situated within the regular regime (labeled 1–3), the FLIs exhibit slow polynomial growth, stably settling at $\log_{10} \text{FLI} \sim 0.5$ by the end of the integration time. In contrast, trajectories within the chaotic sea (labeled 4–6) display rapid exponential FLI growth, reaching $\log_{10} \text{FLI}$$\sim$$2$ over the same timescale. This stark contrast in evolutionary behavior justifies establishing $\log_{10} \text{FLI} > 1.5$ as the criterion for classifying an orbit as chaotic, which is adopted in Fig.~\ref{fig6}.

The right panel of Figure~\ref{figC2} displays the time histories of $e$, $H$, $\sigma_1$, and $\Sigma_2^*$ for trajectory 5 in the Poincar\'e section. The vertical black dashed lines indicate the instants when the trajectory enters the uncertainty zone. It is evident that $\Sigma_2^*$ remains approximately constant outside the uncertainty zone, but undergoes abrupt changes during the passages through this region, which is characterized by extremely high eccentricity and the breakdown of $\sigma_1$ as a slow variable, consistent with the mechanism discussed in Section~\ref{Sect4}.

\section{Dependence of chaotic regions on the strength of GR term}
\label{SectA4}

\begin{figure*}%[ht!]
   \centering
   \includegraphics[width=1.8\columnwidth]{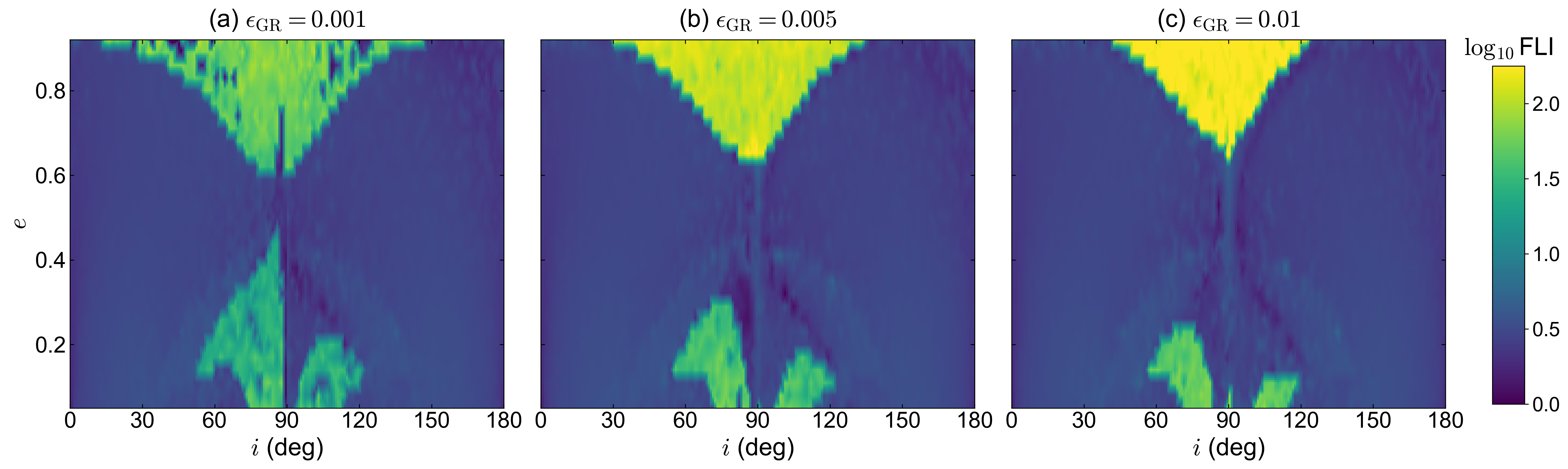}
      \caption{The chaos indicator FLI in the initial eccentricity--inclination plane for different values of $\epsilon_{\text{GR}}$, with initial angle configurations of $(g_0 = 0, h_0 = \pi)$.}
      \label{figD1}
\end{figure*}

To examine how the chaotic region varies with the strength of GR term, we compute the $\log_{10} \text{FLI}$ maps in the $(e_0, i_0)$ plane for several values of $\epsilon_{\text{GR}}$, namely $\epsilon_{\text{GR}} = 0.001, 0.005$, and $0.01$. The results are shown in Fig.~\ref{figD1}.

As $\epsilon_{\text{GR}}$ increases, the chaotic region shrinks progressively, while the FLI values within the remaining chaotic zone reach higher magnitudes over the same integration time. This behavior is consistent with our analytical framework: stronger GR precession shifts the uncertainty zone toward higher inclinations, thereby narrowing the range of initial conditions that can access it. At the same time, trajectories that do enter the zone undergo more rapid changes in the adiabatic invariant, which in turn leads to stronger chaos. This trend is also in agreement with the findings of \citet{liu2015}, who showed that increasing the strength of short-range forces progressively narrows the inclination window for orbital flips.

\section{Comparison between SA and DA trajectories}
\label{SectA5}

The analysis in the main text is based on the double-averaged (DA) Hamiltonian including the Brown correction. Here, we adopt the single-averaging (SA) approximation \citep{huang2026}, which averages the original Hamiltonian over the inner orbit, to clarify the relation between the uncertainty zone identified in this work and the regime where nonadiabatic jumps occur in SA studies \citep{klein2026}.

\begin{figure}
    \centering
    \includegraphics[width=0.95\columnwidth]{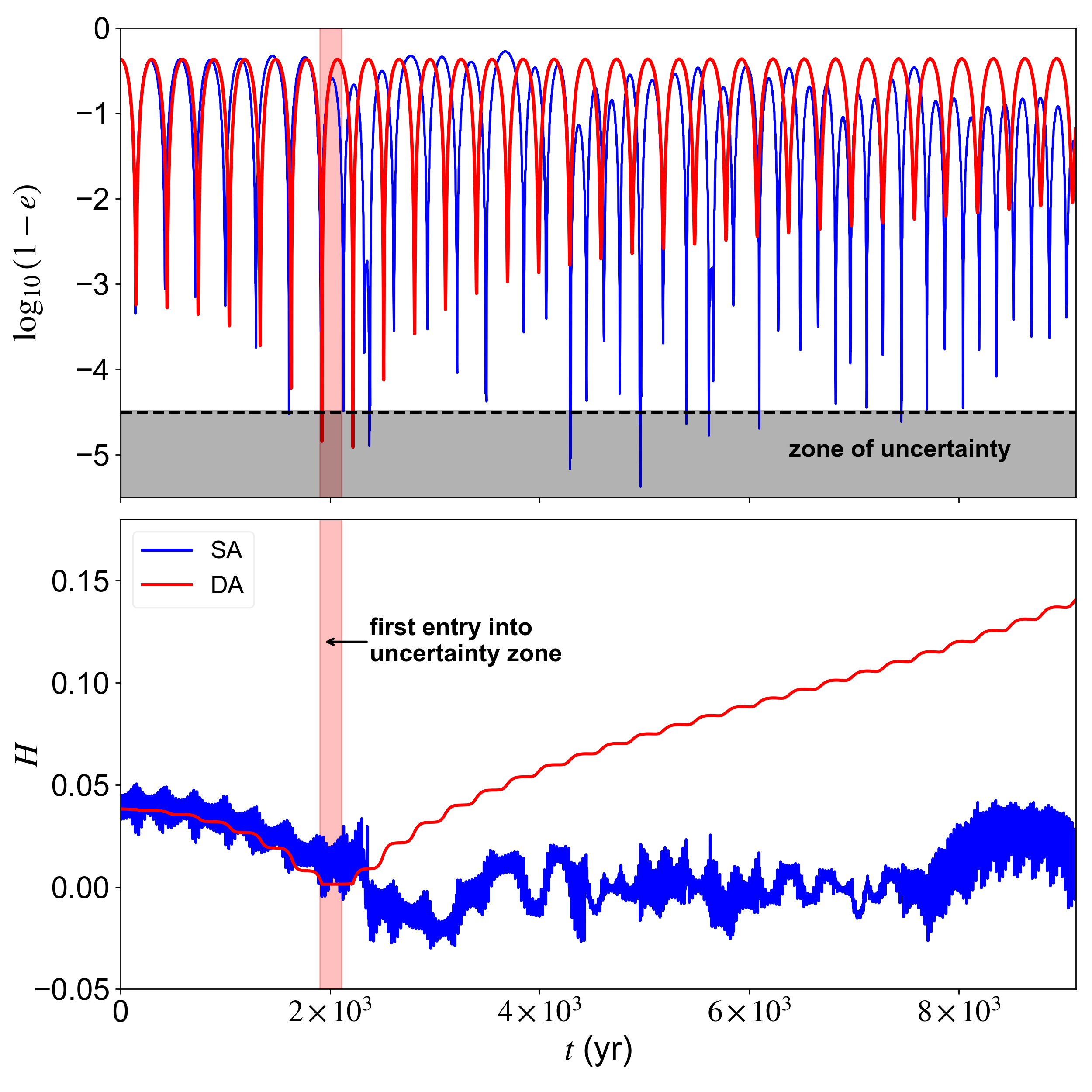}
    \caption{Comparison between SA (blue) and DA (red) trajectories. The gray-shaded horizontal band in the top panel marks the uncertainty zone. As expected, the SA trajectory closely matches the DA one before entering the uncertainty zone and, afterward, they diverge and then the motion becomes chaotic.}

    \label{fig:SA_DA}
\end{figure}

The initial conditions are identical for both SA and DA integrations, with $e_0=0.3,i_0=88^\circ,g_0=0,h_0=0$, and $M_{\mathrm{p},0}=0$. 
A direct comparison of the time histories of eccentricity $e$ and the vertical angular momentum $H=\sqrt{1-e^2}\cos i$ between SA and DA trajectories is presented in Fig.~\ref{fig:SA_DA}, where the blue and red lines denote the SA and DA trajectories, respectively. It is observed that, prior to the first entry into the uncertainty zone, both $e$ and $H$ from the SA trajectory closely follow the DA solution. Once the trajectory enters this zone, however, the two trajectories rapidly diverge and $H$ becomes erratic and irregular, indicating the occurrence of non-adiabatic jumps similar to those identified in \citet{klein2026}. After these jumps have taken place, the subsequent motion becomes chaotic. This is consistent with our conclusion that trajectories traversing the uncertainty zone are intrinsically chaotic.

\end{appendix}

\end{document}